\documentclass[aps,pre,superscriptaddress,twocolumn,showpacs,floatfix]{revtex4}

\usepackage{graphicx}
\usepackage{dcolumn}
\usepackage{bm}

\usepackage{epsfig}
\usepackage{amsmath}

\newcommand{\beq}{\begin{equation}}
\newcommand{\eeq}{\end{equation}}

\newcommand{\bea}{\begin{eqnarray}}
\newcommand{\eea}{\end{eqnarray}}

\begin{document}

\title{Evolution in complex systems:\\ record dynamics in models
 of spin glasses, superconductors and evolutionary ecology.}
\author{Henrik Jeldtoft Jensen}
\email{h.jensen@ic.ac.uk}
\homepage{http://www.ma.ic.ac.uk/~hjjens/}
\affiliation{Department of Mathematics, Imperial College London,
South Kensington campus, London SW7 2AZ, U.K.}


\begin{abstract}
Recent research on the non-stationary nature of the dynamics of complex systems
is reviewed through three specific models. The long time dynamics consists
of a slow, decelerating but spasmodic release of generalized intrinsic strain.
These events are denoted quakes. Between the quakes weak fluctuations occur
but no essential change in properties are induced. The accumulated effect
of the quakes, however, is to induce a direct change in the probability
density functions characterising the system. We discuss how the log-Poisson
statistics of record dynamics may be an effective description of the long 
time evolution and describe how an analysis of the times at which the quakes 
occur enables one to check the applicability of record dynamics. 

\end{abstract}

\maketitle

Out of equilibrium systems are often treated as being in a stationary
state characterised by time independent statistical measures. Although
this is probably the case in some situations there are many instances where
this is not so and where one may miss essential aspects of the behaviour
if attempts are made to treat the phenomena as stationary or nearly stationary.

Complex systems often display evolving macroscopic properties. The most important task
of a theoretical treatment is then to understand the link between the microscopic
fluctuations, which will often exhibit an approximate time reversal symmetry
and the macroscopic directed evolution.  The description should as well
explain the nature of the emergent macroscopic dynamics.

Here we review how the concept of record dynamics, developed by Sibani and
Littlewood\cite{Sibani93a}, has successfully served as a paradigm for the
description of the evolution of three very different models: the relaxation
of a spin glass following an initial temperature quench, the penetration
of an external magnetic field into a disordered type II superconductor and
a model of evolutionary ecology. In all three cases macroscopic variables,
which exhibit a degree of intermittent dynamics, can be identified.
Furthermore, the sequence of transitions between metastable configurations 
can be  analysed in terms of the record statistics. 

The work reviewed here is a result of collaboration with Paolo Sibani,
Paul Anderson and Luis P Oliveria. Some details of the specifics have been published
in \cite{Anderson04,Sibani04,Sibani05,Oliveria05}. The concept of record dynamics
have been developed by Sibani and his collaborators over a long period,
see e.g. \cite{Sibani93a,Sibani99a,Dall01,Dall03,Sibani03}.

Below we first introduce the models in sufficient self-contained detail.
Next we describe how the long time dynamics in each case are manifestations
of record dynamics and discuss its consequences.

\section{Three models}
Here follows a brief description of the definition of the microscopic dynamics
of the three models considered.
 
\subsection{Spin glass}
\label{Spin-glass-A}
We consider a three dimensional Edwards-Anderson spin glass
\begin{equation}
H=-\frac{1}{2}\sum_{\langle ij\rangle } J_{ij} S_iS_j.
\label{spin-glass}
\end{equation}
with nearest neighbour Gaussian couplings\cite{Edwards75} 
and Ising spin $S_i=\pm1$. At time zero the temperature is instantaneously
dropped from infinity to a very low value. The subsequent dynamics is realised
by use of Monte Carlo dynamics, see \cite{Sibani04,Sibani05,Dall03}.

\subsection{Magnetic relaxation}
\label{ROM-A}
We use Monte Carlo (MC) simulations of a generalized  
three dimensional layered version of the Restricted Occupancy Model (ROM) model  
to capture the long time relaxation of interacting vortex matter
\cite{nicodemi_1,nicodemi_2,nicodemi_3,nicodemi_4,nicodemi_5,nicodemi_6,Oliveria05}.

The length scales of vortex interactions can be very large
compared with the average separation between vortices. At high magnetic
induction each vortex interacts with many others suggesting 
that a simplified coarse grained description in terms of vortex densities may
be applicable.  For layered superconductors it is natural to introduce
two separate length scales: the first is the range of the interaction parallel to
the planes, this is the London
penetration depth $\lambda$. The second length
scale is the vortex correlation length, $\xi_{||}$, parallel to the
applied field (which we imagine to be perpendicular to the   
copper oxide planes for high temperature superconductors).
The exact identification of this length scale is difficult
and is likely to depend on the anisotropy of the material,
the nature of the pinning, the strength of the magnetic induction
and on the temperature. This length scale may be related to vortex
line 
cutting\cite{Puig00,Goffman98,delaCruzBSCO,HenrikYBC0,OdeficientYBC0,Gaifullin,Busch,Fuchs}.  
These length scales  respectively give the horizontal, $l_0$,
and vertical, $l_1$, coarse-graining length and therefore the lattice spacing of our model. 
Horizontally we have $l_0=\lambda$ and perpendicularly $l_1\sim \xi_{||}$. 
Smaller length scales are ignored. 
For our purposes this approximation is acceptable because the length 
scales smaller than $\lambda$ seem to have little influence on the 
long time glassy properties of vortex matter.

The behaviour of vortex matter is
determined by the competition of four energy scales \cite{Blatter}: intra
and interlayer vortex-vortex interaction, vortex-pinning interaction and
thermal fluctuations, all of which are schematically included in the ROM
model.\bigskip

The Hamiltonian of the ROM model is thus the following:
\begin{equation}
H=\sum_{ij}A_{ij}n_{i}n_{j}- \sum_{i}A_{ii}n_{i}+\sum_{i}A_{i}^{p} n_{i} +
\sum_{\left\langle ij\right\rangle _{z}}A_{2}\left(  n_{i}-n_{j}\right)  ^{2},
\label{hamilton}
\end{equation}
where $n_{i}$ is the number of vortices on site $i$ of the lattice. 
In a superconducting sample the number of vortex lines per unit area is
restricted by the upper critical field ($B_{c2}$) \cite{tinkham}, so in the
model the number of vortices per cell can only assume values smaller than
$N_{c2}=B_{c2}l_{0}^{2}/\phi_{0}$ \cite{Nicodemi_and_Henrik_c1,nicodemi_4}.
Hence the name Restricted Occupancy Model. Moreover,
as we are interested in
a simulation setup that does not require magnetic field inversion and the
vortex-antivortex creation is strongly suppressed, 
we simply consider $n_{i}\geq0$.

The first two terms in Eq. (\ref{hamilton}) 
represent the repulsion energy due to vortex-vortex
interaction in the same layer, and the vortex self energy respectively. Since 
the potential that mediates this interaction decays exponentially at
distances longer than our coarse-graining length $\lambda$, interactions 
beyond nearest neighbours are neglected. We set
$A_{ii}:=A_{0}=1$, $A_{ij}:=A_{1}$ if $i$ and $j$ are nearest neighbours on
the same layer, and $A_{ij}:=0$ otherwise.

The third term represents the interaction of the vortex pancakes with the pinning
centres. $A_i^{p}$ is a random potential and for simplicity we
consider that $A_i^{p}$ has the following distribution $P\left(  A_i^{p}\right)
=\left(  1-p\right)  \delta\left(  A_i^{p}\right)  -p\delta\left(  A_i^{p}
-A_{0}^{p}\right)$.
The pinning strength $\left|  A_{0}^{p}\right|$ represents the total
action of the pinning centres located on a site. In the present work
we use $\left|  A_{0}^{p}\right|  =0.3$.

Finally the last term describes the interactions between the vortex sections
in different layers. This term is a nearest neighbour quadratic interaction
along the $z$ axis, so that the number of vortices in neighbouring cells along
the $z$ direction tends to be the same.

The parameters of the model are defined in units of $A_{0}$. The
time is measured in units of full MC 
sweeps. The relationship between the model parameters and material parameters is
discussed in \cite{Nicodemi_and_Henrik_c1,nicodemi_4}.
The model has been demonstrated to reproduce a very broad range of experimental 
observations including dynamical aspects of magnetic creep and memory
and rejuvenation of voltage-current 
characteristics\cite{jackson,nicodemi_1,nicodemi_2,nicodemi_3,nicodemi_4,nicodemi_5,nicodemi_6}.  

Each individual MC update involves the movement to a neighbour site
of a single randomly selected vortex. The movement of the vortex is
automatically accepted if the energy of the system decreases; if the energy of
the system increases, the movement is accepted with probability 
$\exp(-\Delta E/T)$ \cite{Binder}. 

The external magnetic field is modelled by the edge sites on each
of the planes. The density at the edge is kept at a controlled value. 
During a MC sweep vortices may move between the bulk sites and the edge sites.
After each MC sweep the density on the edge sites is brought 
to the desired value. Initially the external field is increased to
a desired value ($N_{ext}=10$ vortices per edge site) by a very rapid 
increase in the  density on the edge sites.  

After this fast initial ramping the external field is kept constant, while
we study how the vortices move into the sample. The age of the system, $t_w$,
is taken to be the time since the initial ramping. 

\subsection{Tangled Nature}
\label{Tangled-Nature-A}
\subsection{\textit{Definition of the model}}
The Tangled Nature model is an individual based model of evolutionary
ecology. We give a brief outline of the model here. 
Details can be found in \cite{christensen02,Hall02,Anderson04}.
An individual is represented by a vector ${\bf S}^\alpha=
(S_1^\alpha,S_2^\alpha,...,S_L^\alpha)$ in the genotype space
$\cal{S}$, where the ``genes'' $S^\alpha_i$ may take the values $\pm 1$, 
i.e. ${\bf S}^\alpha$ denotes a corner of the 
$L$-dimensional hypercube. In the present paper we take $L=20$ as this
gives space of a reasonable size to explore (over a million genotypes)
whilst not being computationally prohibitive. 
We think of the genotype space $\cal{S}$ as containing all possible ways
of combining the genes
into genome sequences. Many sequences may not correspond to viable
organisms. The viability of a genotype is determined by the
evolutionary dynamics. All possible sequences are made available for evolution to select from.
The number of occupied sites is referred to as the diversity, here
analogous to the number of species or species richness
\cite{kreb:ecol}. As explained later, genotype, species, site and node
are synonymous throughout.

For simplicity, an individual is removed from the system with a
constant probability $p_{kill}$ per time step.   
A time step consists of {\em one} annihilation attempt followed by
{\em one} reproduction attempt. One generation 
consists of $N(t)/p_{kill}$ time steps, which is the average time 
taken to kill all currently living individuals. All references to time
will be in units of generational time.

The ability of an individual to reproduce is controlled by a weight
function $H({\bf S}^\alpha,t)$:  
\begin{equation}
H({\bf S}^\alpha,t)={1\over c N(t)} \left( \sum_{{\bf S}\in{\cal S}} 
J({\bf S}^\alpha,{\bf S})   n({\bf S},t) \right)
- \mu N(t),
\label{Hamilton2} 
\end{equation}
where $c$ is a control parameter, $N(t)$ is  the total number of individuals at time $t$, 
the sum is over the $2^L$ locations in ${\cal S}$ and $n({\bf S},t)$ is 
the number of individuals (or occupancy) at position ${\bf S}$. 
Two positions ${\bf S}^a$ and ${\bf S}^b$ in genome space are coupled with the fixed random 
strength $J^{ab}=J({\bf S}^a,{\bf S}^b)$ which can be either positive, negative or zero. 
This link is non-zero with probability $\theta$, i.e. $\theta$ is
simply the probability that any two sites are interacting. To study the
effects of interactions \textit{between} species, we exclude
self-interaction so that $J^{aa}=0$. 

The conditions of the physical environment are simplistically
described by the term $\mu N(t)$ in equation (\ref{Hamilton2}), 
where $\mu$ determines the average sustainable total population size, i.e. 
the carrying capacity of 
the environment. An increase in $\mu$ corresponds to harsher physical
conditions. Notice that genotypes only adapt to each other and the
physical environment represented by $\mu$. 
We use asexual reproduction consisting of one individual
being replaced by two copies mimicking the process of binary fission
seen in bacteria.
Successful reproduction occurs with a probability per unit time given by
\begin{equation}
p_{off}({\bf S}^\alpha,t)={ \exp[H({\bf S}^\alpha,t)]\over
1+\exp[H({\bf S}^\alpha,t)]}\in[0,1].
\label{p_off}
\end{equation}
This function is chosen for convenience. We simply need a smoothly
varying function that maps $H({\bf S}^\alpha,t)$ to the interval
$[0,1]$ and it is otherwise arbitrary.
We allow for mutations in the following way: with probability $p_{mut}$ per gene we perform a 
change of sign $S_i^\alpha \rightarrow - S_i^\alpha$ during reproduction. 

Initially, we place $N(0)=500$ individuals at randomly chosen positions. 
Their initial location in genotype space does not affect the nature of the
dynamics. A two-phase switching dynamic is seen consisting of long
periods of relatively stable configurations (quasi-Evolutionary Stable Strategies
or q-ESSs) interrupted by brief spells of reorganisation of occupancy
which are terminated when a new q-ESS is found, as discussed in \cite{christensen02}.

\section{Record dynamics and its manifestation}
In this section we review the macroscopic intermittent dynamics of the
three models and show that in all cases record dynamics is an efficient
description of the statistical aspects of the temporal evolution. Before
that we need to sketch the notion of {\it record statistics}. 

Let $\chi (t)$ denote an uncorrelated stochastic signal distributed according
to the probability density function(pdf) $p(\chi)$. By the record of the
signal we mean $R(t)={\rm max}\{\chi(t')|t'\leq t\}$. Obviously $R(t)$
is a piecewise constant function which jumps discontinuously as a fluctuation 
manages to take $\chi(t)$ to a new record value. The times $t_k$ at which this happens
are called the record times.
It was pointed out be Sibani and Littlewood\cite{Sibani93} that the probability
that exactly $q$ records occure 
during a time interval $[t_w,t_w+t]$ is to a good approximation given by
\begin{equation}
p(q)=\frac{\langle q\rangle^q}{q!}\exp \{-\langle q\rangle\},
\label{log-P}
\end{equation}
where $\langle q\rangle = \alpha \log(1+t/t_w)$. This is a Poisson distribution
in the logarithm of time. For a mathematical process the logarithmic rate $\alpha=1$.
Here we include the possibility $\alpha \neq 1$, which may happen for a
physical process as an effect of over or undercounting of the true number
of records. For example $\alpha>1$ can occur if the recorded record times are
produced by more than a single independent record process. In contrast
$\alpha<1$ can e.g. be an effect of not being able resolve all the record
events of record process. 
The average number of records per time unit decreases inversely
proportional with time, namely 
\begin{equation}
\frac{d \langle q\rangle}{dt}=\frac{\alpha}{t_w+t}.
\label{insta_acti}
\end{equation}
It is important to note that Eqs. (\ref{log-P}) and (\ref{insta_acti})
are independent of $p(\chi)$. Thus the statistical properties of a record
signal $R(t)$ are very general and do not depend on the properties of the 
underlying fluctuating signal $\chi(t)$.

\subsection{Spin-glass}
\label{spin-glass-B}
After an initial quench from very high temperature it is natural that the
dynamics of the spin glass leads to a relaxation towards ever lower energy.
The specifics of how this happens was analysed in great detail by Sibani and 
collaborators\cite{Sibani03,Dall03} for the Edwards-Anderson spin glass.
They followed the temporal evolution of the total energy $E(t)$ given in
Eq. (\ref{spin-glass}. They identified the sequence of local minima $E_{min}(k)$
and local maxima $E_{max}(k)$ from which they defined the $k$-th barrier
as $\Delta E_k=E_max(k)-E_min(k)$. The set of barriers turn out to be monotonously
increasing $\Delta E_k<\Delta E_{k+1}$, but only marginally so in the
sense that $\Delta E_k\simeq \Delta E_{k+1}$. Thus to exit the $k$-th metastable 
state visited by the spin glass, a barrier slightly larger than any encountered
previously has to be overcome. The set of time instances at which the spin glass manage
to move from one metastable configuration, i.e. the quake times $t_k$,
was found to follow the  log-Poisson distributed characteristic of record statistics,
see E.q. (\ref{log-P}). 

\subsection{Magnetic relaxation}
\label{ROM-B}
\begin{figure}
\vspace{.3cm}
\begin{center}
\includegraphics[width=8cm]{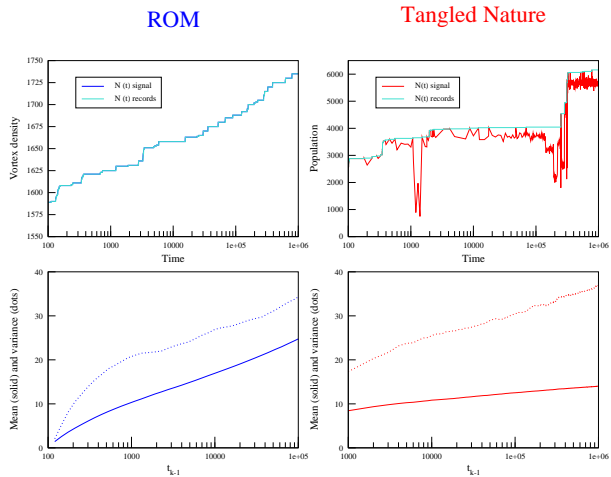}

\end{center}
\caption{Time dependence of the ROM and the Tangled Nature model}
\label{N(t)}
\end{figure}

The magnetic pressure exerted by the external magnetic field in the ROM
model introduced above will force the number of vortices in the bulk, $N_v(t)$, of
the sample to increase with time. We have found that $N_v(t)$ is essentially
a record signal\cite{Anderson04,Oliveria05}, see Fig. \ref{N(t)}.

The sequence of record times follow to good approximation the statistics
derivable from Eq. (\ref{log-P}) expected for an uncorrelated record signal\cite{Anderson04}.

\subsection{Evolutionary Ecology}
\label{Tangled-Nature-B}
That directedness of the temporal evolution of the quenched spin glass
and also of the superconductor in an external field is to be expected.
It is less obvious why the Tangled Nature  model of biological evolution exhibit
a gradual adaptation towards more stable configurations. The diffusive nature
of the dynamics in genotype space may suggest the breaking of time reversal
symmetry. But this in itself doesn't point to a reason why the total number
of individuals in the model is increasing, on average, with time. Nor does
diffusion imply that the configurations tend to be come more stable with
time. 

The following analysis suggests an explanation. First assume that no mutations
can occur. The population dynamics is controlled by the fixed probability
$p_{kill}$ and the offspring probability $p_{off}$. Fluctuations in the
population size $N(t)$ will lead to fluctuations in $p_{off}$ according
to Eq. (\ref{Hamilton2}). The overall stability
of the population is ensured since the fixpoint condition $p_{off}=p_{kill}$
is stable. This follows because an increase in $N(t)$ will lead to a decrease in $H$
(see Eq. (\ref{Hamilton2}) and therefore in $p_{off}$. Similarly  a decrease in 
$N(t)$ leads to a increase in $H$. Now consider the effect of mutations.
The evolutionary dynamics is driven by the mutations which move individuals
between positions in genotype space. These mutations are random and
lead to symmetric fluctuations in the weight function $H$ 
as an effect of changing the coupling term $H$, see Eq. (\ref{Hamilton2}).
Let us schematically write
$H$ as $H={\cal J}-\mu N$. Mutations induce fluctuations of the form ${\cal
J} \mapsto {\cal J}+\delta {\cal J}$. Assume $\delta {\cal J}>0$ and
that the same fluctuation with opposite sign $-\delta {\cal J}$ occurs
with equal probability. The mutation leading to $\delta {\cal J}$ is, however,
more likely to become established since 
\begin{equation}
p_{off}(h+\delta {\cal J})>p_{off}(H-\delta {\cal J}),
\end{equation}
for values of $H$ where  $p_{off}$ is convex; which is the case for $p_{off}<1/2$
according to Eq. (\ref{p_off}). This is a mathematical way of
paraphrasing Darwin's description of how favourable variations 
become entrenched\cite{Darwin}.

It was found from simulations of the Tangled Nature model that $N(t)$ gradually
increases. Furthermore, the configurations occupied in genotype space
becomes, on average, more stable with time\cite{Hall02,Anderson04}. In
Fig. \ref{N(t)} we compare the record signal derive from $N(t)$ with $N(t)$.
It is clear that there are much larger deviations between the record and
the signal itself, than was the case in the ROM model. Nevertheless, we do
believe that the statistics of the record times obtained from the record
signal derive from $N(t)$  essentially corresponds to the transition times
between the q-ESS epochs.

\section{Consequences}
We briefly review some of the most prominent consequences, which may be derived from the
record dynamics.

\subsection{Spin glasses}
\label{Spin-glass-C}
A detailed description of how certain aspect of intermittency, aging  and memory 
in spin glasses can be considered a consequence of record dynamics was given
in \cite{Sibani04,Sibani05}. That the fluctuations clearly separates
into `quake' fluctuations and Gaussian equilibrium like fluctuations show
up directly in a study of the heat exchange between the spin glass and
its thermal bath. The pdf for the heat exchange consists of a  Gaussian
part
and an exponential tail. The exponential tail is produced by the large
releases of energy that occur when a quake takes the spin glass from one metastable
state to the next. The exponential tail is only visible when the heat exchange
is collected during a time interval, $\delta t$ that is short compare to the time  
$t_w$ passed since the initial quench from high temperature. Recall that
the quake activity decreases roughly inversely  proportional with time. If $\delta t\gg t_w$
there is not enough quake activity during the sampling time $\delta$ to make
the exponential tail of the pdf visible though peak of the much more frequent Gaussian
fluctuations. In this way the one point distribution of the heat exchange is able
to probe the aging of the spin glass\cite{Sibani05}.

Aspects of memory and rejuvenation in a temperature shift experiment can also be 
related to the record dynamics\cite{Sibani04}. Let $b(t_w,T)$ denote the
largest energy barrier overcome by the quakes during the time $t_w$ since
the initial quench. This barrier will, according to the Arrhenius law,
determine the rate of quake activity, $r_q$ at times about $t_w$. On the other
hand, from the view point of record dynamics, we also have
$r_q(T)\propto1/t_w$. A drop in the temperature $T\mapsto T'$ at $t_w$ will not change
the barrier $b(t_w, T)$ established by the record dynamics prior to $t_w$.
The Arrhenius activation of the quakes will however drop when the temperature
is lowered producing an effective age of the system $t_w^{eff}\propto1/r_q(T')>t_w$.
It is interesting to note that the effect of the temperature drop has two
opposite effects concerning the apparent age of the system: (a) the
drop in $T$ make the energy barriers look bigger. So in this respect the
spin glass appears to be older. (b) The amount of energy delivered to the
heat bath during a quake is higher and in this respect the spin glass appears
younger. A detailed discussion is given in Ref. \cite{Sibani04}.

\subsection{Magnetic relaxation} 
\label{ROM-C}
\begin{figure}
\vspace{.3cm}
\begin{center}
\includegraphics[width=8cm]{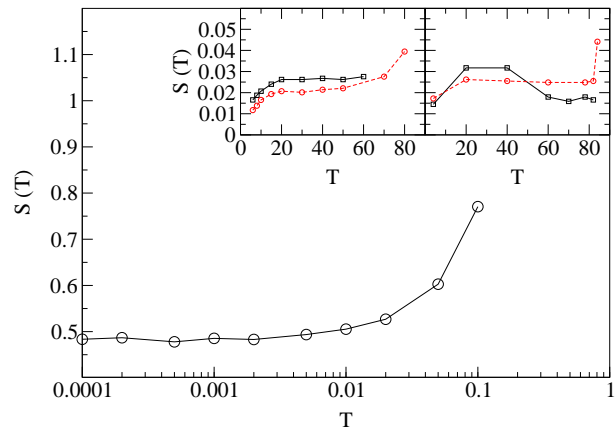}

\end{center}
\caption{Main panel: Numerical results for the creep rate
versus temperature. Details in \cite{Oliveria05}. 
Insets:
experimental
results for the creep rate versus T.
The right inset shows data from Keller {\it et al.} for melt
processed
YBCO crystals with the magnetic field applied along the $c$ axis (squares)
and $ab$ plane (circles). The left inset shows data from
\cite{Civale_et_al.} for unirradiated (squares) and 3 MeV proton-irradiated
(circles) YBCO
flux grown \cite{Kaiser_et_al.} crystals with a 1 T magnetic Field applied
parallel to the $c$ axis.} 
\label{creep_rate}
\end{figure}

We now explain how record dynamics might explain the observation that
the rate of thermally activated creep is found to be essentially temperature
independent for broad ranges of the temperature\cite{Oliveria05}. We described
in section \ref{ROM-A} and \ref{ROM-B} that, in the ROM model, the temporal evolution of the
total number of vortices in the bulk of the sample is a record signal.
A prominent feature of the record time of an uncorrelated stochastic process
is that the statistics of the records, and in particular the rate with which
the records occur, is independent of the properties of the underlying stochastic process.
For the specific case of thermally induced fluctuations this implies that the rate of the record
will be temperature independent. In Fig. \ref{creep_rate} we show that
the temperature dependence of the ROM model compare well with experimental
creep rates. Both exhibit a broad range of weak temperature dependence.
Details can be found in \cite{Oliveria05}

\subsection{Biological evolution}
\label{Tangled-Nature-C}
In figure \ref{N(t)} we showed that the total population size of the
Tangled Nature model, despite of large fluctuations, may also be related
to record dynamics. The nature of the metastable states between the quakes
is in this case difficult to determine. But we believe they are closely
related to the quasi-Evolutionary Stable Strategies which have been identified
in the model\cite{christensen02}. We have found that the number of extinctions
and creation events per time decreases in the Tangled Nature model\cite{Hall02}.
Similar behaviour is observed in macro-evolution\cite{Newman99c}. The dynamics
of the Tangled Nature model is intermittent as is to some extend clear from
Fig. \ref{N(t)}. The intermittency is, however, much more evident from
analysis of the time dependence of the configurations in genotype, see 
space\cite{christensen02,Hall02}. The fossil record has also been 
interpreted as exhibiting intermittency, see e.g. \cite{eldredge_gould}.
Accordingly by comparing the Tangled Nature model and the dynamics of the
fossil record we are able to suggest that the decreasing  extinction rate
and the intermittency, or punctuated equilibrium, is a result a hitherto
unrevealed record dynamic that somehow  controls the macro-dynamics of biological evolution.

\section{Summary and Conclusions}
The relevance of record dynamics to the long time evolution of complex
systems was indicated by reviewing three very different model studies.
The most prominent characteristics of record dynamics are log-Poisson 
distribution of the number of records, or quakes, occurring during a time interval 
and  a rate of events, which decreases inversely proportional with time. 
It will be very interesting to analyse the dynamics of other complex systems 
from the view point of record statistics. The analysis requires
access to the time instances at which quakes occur. If that is not available analysis
of the probability density function of a single fluctuating quantity such as the heat exchange of
a spin glass may suffice to reveal the existence of an underlying record dynamics.



\end{document}